# Emerging applications of integrated optical micro-combs for analogue RF and microwave photonic signal processing

Xingyuan Xu,[1] Jiayang Wu,[1] Sai T. Chu,[3] Brent E. Little,[4] Roberto Morandotti,[5] Arnan Mitchell,[2] and David J. Moss[1,*]

[1]*Centre for Micro-Photonics, Swinburne University of Technology, Hawthorn, VIC 3122 Australia*
[2]*School of Engineering, RMIT University, Melbourne, VIC 3000, Australia*
[3]*Department of Physics and Material Science, City University of Hong Kong, Hong Kong, China.*
[4]*Xi'an Institute of Optics and Precision Mechanics Precision Mechanics of CAS, Xi'an, China.*
[5]*INRS –Énergie, Matériaux et Télécommunications, Varennes, Québec, J3X 1S2, Canada.*
*\*dmoss@swin.edu.au*

## Abstract

We review new applications of integrated micro-combs in RF and microwave photonic systems. We demonstrate a wide range of powerful functions including a photonic intensity high order and fractional differentiators, optical true time delays, advanced filters, RF channelizer and other functions, based on a Kerr optical comb generated by a compact integrated micro-ring resonator, or "micro-comb". The micro-comb is CMOS-compatible and contains a large number of comb lines, which can serve as a high-performance multi-wavelength source for the transversal filter, thus greatly reduce the cost, size, and complexity of the system. The operation principle of these functions is theoretically analyzed, and experimental demonstrations are presented.

## 1. INTRODUCTION

Nonlinear optics for all-optical signal processing has proven to be an extremely powerful approach, particularly when implemented in photonic integrated circuits based on highly nonlinear materials such as silicon [1, 2]. Functions that have been demonstrated include all-optical logic [3], demultiplexing from 160Gb/s [4] to over 1Tb/s [5], to optical performance monitoring using slow light at speeds of up to 640Gb/s [6-7], all-optical regeneration [8,9], and many others [10-16]. CMOS compatible platforms are centrosymmetric, and so nonlinear devices have been based on a range of third order nonlinear processes including third harmonic generation [11,17-21] and, more commonly, the Kerr nonlinearity ($n_2$) [1,2]. The efficiency of Kerr based all-optical devices depends on the waveguide nonlinear parameter, $\gamma = \omega n_2 / c A_{eff}$ where $n_2$ is the Kerr nonlinearity and $A_{eff}$ is the waveguide effective area). Although silicon can achieve extremely high values of $\gamma$, it suffers from high nonlinear losses due to two-photon absorption (TPA) and the resulting free carriers [2]. Even if the free carriers are eliminated by p-i-n junctions, silicon's poor intrinsic nonlinear figure of merit (FOM = $n_2 / (\beta \lambda)$, where $\beta$ is the TPA) of around 0.3 in the telecom band is far too low to achieve high performance. While TPA can be turned to advantage for some all-optical functions [22-24], for the most part silicon's low FOM in the telecom band is a limitation. This has motivated research on a range of alternate nonlinear platforms highlighted by perhaps by the chalcogenide glasses[25-35]. While offering many advantages, these platforms however do not offer the compatibility with the silicon computer chip industry – CMOS (complementary metal oxide semiconductor).

In 2008 to 2010, new CMOS compatible platforms for nonlinear optics, including silicon nitride [35,36] and Hydex [37-47] were introduced that exhibit negligible nonlinear absorption in the telecom band, a moderate nonlinear parameter and extremely high nonlinear figure of merit, which are ideal for micro-comb generation [47]. Following the first report of Kerr frequency comb sources in 2007 [48], the first integrated CMOS compatible integrated optical parametric oscillators were reported in 2010 [46, 47], and since then this field has exploded [47]. Many cutting-edge applications have been demonstrated based on CMOS-compatible micro-combs, ranging from filter-driven mode-locked lasers [49-52] to quantum physics [53-58]. Most recently, Kerr micro-combs have demonstrated their enormous potential for sources for ultrahigh bandwidth coherent optical fiber communications[59], optical frequency synthesis [60] and many other applications [61-68]. The success of these new CMOS platforms as well as other CMOS platforms such as amorphous silicon [69] arises from a combination of their low linear loss, moderate to high nonlinearity and low or even negligible nonlinear loss (TPA).

An emerging and very powerful application of micro-ring resonator based Kerr frequency combs - "microcombs" [47] - has been in the area of radio frequency (RF) and microwave photonics that bring together the worlds of radiofrequency engineering and optoelectronics [70-90]. RF and microwave photonics exploit the potential of optical technologies and offers many benefits for RF systems including high speed, broad operation bandwidth, low loss, and strong immunity to electromagnetic interference. A diverse range of photonic approaches to RF signal generation, transmission, processing, and sensing have been proposed and widely employed in RF systems and communication networks[91-117]. However, most RF systems are composed of discrete components, which impose certain drawbacks in terms of cost, power consumption and reliability, thus holding RF photonic systems from reaching maturity and replacing traditional RF solutions. As one of the most powerful tools in RF photonic systems, optical frequency combs offer an alternative as multi-wavelength sources for multiple RF channels, and thus can greatly increase the capacity for transmission and performance for transversal processors. This is particularly true of optical micro-combs [47, 118-129].

Here, we review our recent progress in applications of optical micro-combs to RF and microwave photonics [118]. In particular, we demonstrate a reconfigurable RF photonic intensity differentiator [124] based on CMOS-compatible micro-combs. By employing an on-chip nonlinear micro-ring resonator (MRR), we generate a broadband Kerr comb with a large number of comb lines and use it as a high-quality multi-wavelength source for a transversal differentiator. The large frequency spacing of the integrated Kerr comb yields a potential operation bandwidth of over 100 GHz, well beyond the processing bandwidth of electronic devices. By programming and shaping the power of individual comb lines according to corresponding tap weights, reconfigurable intensity differentiators with variable differentiation orders can be achieved. Detailed analyses of the operation principle and experimental demonstrations of fractional-, first-, second-, and third-order intensity differentiations are performed. We also present results on optical true time delays for radar applications [129], programmable RF filters [122] and other advanced functions [122]. Optical micro-combs offer powerful solutions to implement RF and microwave photonic signal processing functions.

## 2. OPERATION PRINCIPLES

Based on the theory of signals and systems, the spectral transfer function of an *N*-th order temporal differentiator can be expressed as

$$H(\omega) \propto (j\omega)^N, \quad (1)$$

where $j = \sqrt{-1}$, $\omega$ is the angular frequency, and $N$ is the differentiation order. According to the above transfer function, the amplitude response of a temporal differentiator is proportional to $|\omega|^N$, while the phase response has a linear profile, with a zero and $\pi$ jump at zero frequency for $N$ even and odd, respectively. The ideal RF amplitude and phase responses of first-, second-, and third-order microwave differentiators are shown in Figs. 1(a)–(c), respectively.

We employ [114] a versatile approach towards the implementation of microwave photonic differentiators based on transversal filters, where a finite set of delayed and weighted replicas of the input RF signal are produced in the optical domain and combined upon detection. The transfer function of a typical transversal filter can be described as

$$H(\omega) = \sum_{n=0}^{M-1} a_n e^{-j\omega nT}, \quad (2)$$

where $M$ is the number of taps, $a_n$ is the tap coefficient of the $n^{th}$ tap, and $T$ is the time delay between adjacent taps. It should be noted that the differentiator designed based on Eq. (2) is an intensity differentiator, i.e., the output RF signal after being combined upon detection yields an exact differentiation of the input RF signal, in contrast to field differentiators that yield the derivative of a complex optical field [92-96]. Figure 1(f) depicts the difference between field differentiation and intensity differentiation.

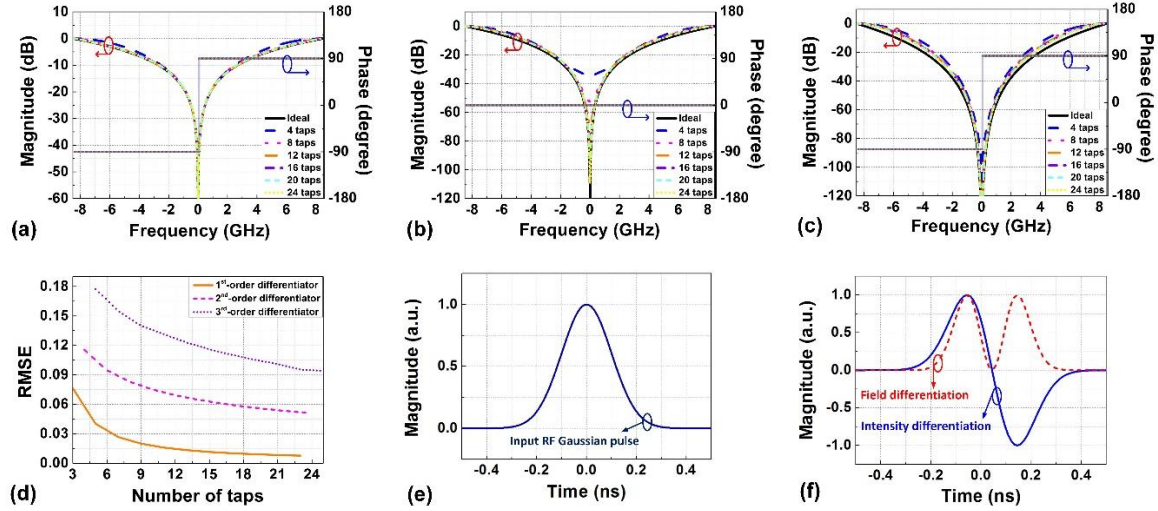

Fig. 1. Simulated RF amplitude and phase responses of the (a) first-, (b) second-, and (c) third-order temporal differentiators. The amplitude and phase responses of the differentiators designed based on Eq. (2) with different number of taps are also shown accordingly. (d) RMSEs between calculated and ideal RF amplitude responses of the first-, second-, and third-order intensity differentiators as a function of the number of taps. Simulated (e) input Gaussian pulse and its corresponding first-order differentiation results.

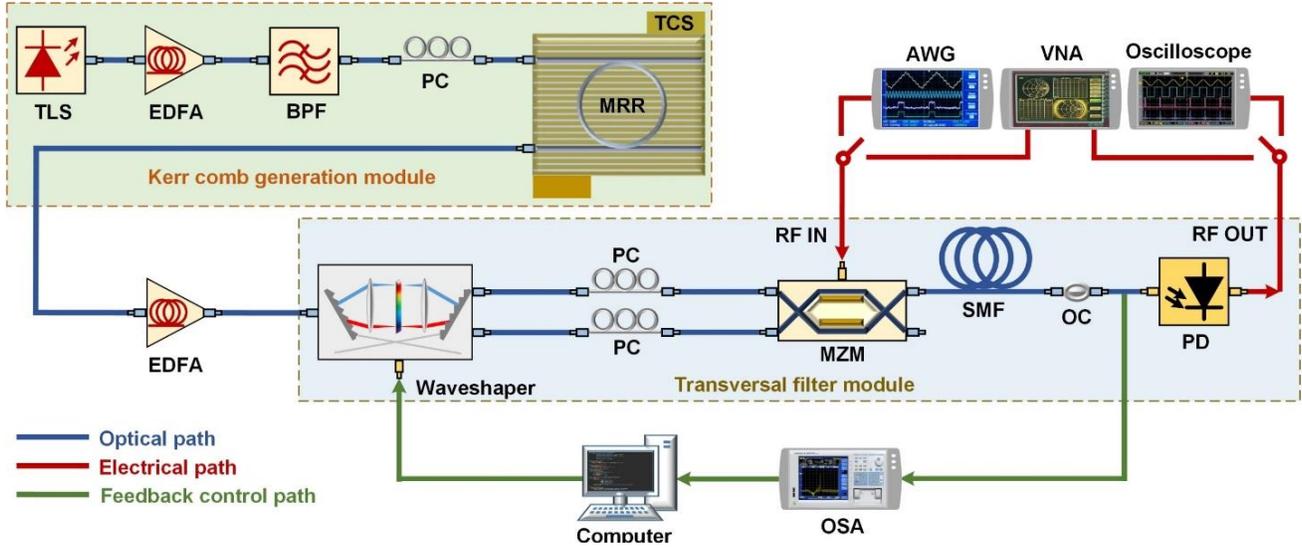

Fig. 2. Schematic illustration of the proposed reconfigurable microwave photonic intensity differentiator. TLS: tunable laser source. EDFA: erbium-doped fibre amplifier. PC: polarization controller. BPF: optical bandpass filter. TCS: temperature control stage. MZM: Mach-Zehnder modulator. SMF: single mode fibre. OC: optical coupler. PD: photo-detector. OSA: optical spectrum analyzer. VNA: vector network analyzer. AWG: arbitrary waveform generator.

To implement the temporal differentiator in Eq. (1), we calculate the tap coefficients in Eq. (2) based on the Remez algorithm [114]. The corresponding amplitude and phase responses of the first-, second-, and third-order differentiators as a function of the numbers of taps are also plot in Figs. 1(a)–(c). When the number of taps is increased, it is clear that the discrepancies between the amplitude response of the transversal filters and the ideal differentiators are improved for all three orders, whereas the phase response of the transversal filters is identical to that of the ideal differentiators regardless of the number of taps. To quantitatively analyze the discrepancies in the amplitude response, we further calculate the root mean square errors (RMSEs) for the first-, second-, and third-order differentiators as a function of the number of taps, as shown in Fig. 1(d). One can see that the RMSE is inversely proportional to the number of taps, as reasonably expected. In particular, we note that when the number of taps increases, the RMSE decreases dramatically for a small number of taps, and then decreases more gradually as the number of taps becomes larger.

Figure 2 shows a schematic illustration of the reconfigurable microwave photonic intensity differentiator [124]. It consists of two main blocks: one is a Kerr comb generation module based on a nonlinear MRR and the other is a transversal filter module for reconfigurable intensity differentiation. In the first module, the continuous-wave (CW) light from a tunable laser source is amplified by an erbium-doped fibre amplifier (EDFA), followed by a tunable optical bandpass filter to suppress the amplified spontaneous emission noise. A polarization controller is inserted before the nonlinear MRR to make sure that the polarization state matches the desired coupled mode. When the wavelength of the CW light is tuned to a resonance of the nonlinear MRR and the pump power is high enough for sufficient parametric gain, the optical parametric oscillation (OPO) process in the nonlinear MRR is initiated, which generates a Kerr optical comb with nearly equal line spacing. The nonlinear MRR is mounted on a highly precise temperature control stage to avoid resonance drifts and maintain the wavelength alignment of the resonance to the CW light. Owing to the compact size and ultra-high quality factor of the nonlinear MRR, the generated Kerr comb provides a large number of wavelength channels with narrow linewidths for the subsequent transversal filter module. As compared to conventional intensity differentiators based on laser diode arrays, the cost, size and complexity can be greatly reduced. After being amplified by another EDFA, the generated Kerr comb then is directed to the second module. The Kerr comb is processed by a waveshaper to get weighted taps according to the coefficients calculated by means of the Remez algorithm. Considering that the generated Kerr comb is not flat, a real-time feedback control path is introduced to read and shape the comb lines' power accurately. A 2×2 balanced Mach-Zehnder modulator (MZM) is employed to generate replicas of the input RF signal. When the MZM is quadrature-biased, it can simultaneously modulate the input RF signal on both positive and negative slopes, thus yielding modulated signals with opposite phases and tap coefficients with opposite algebraic signs. After being modulated, the tapped signals from one output of the MZM are delayed by a dispersive fibre. The time delay between adjacent taps is determined jointly by the frequency spacing of the employed comb source and the dispersion accumulated in the fibre. Finally, the weighted and delayed taps are combined upon detection and converted back into RF signals to form the differentiation output.

Due to the intrinsic advantages of transversal filters, this scheme features a high degree of reconfigurability in terms of processing functions and operation bandwidth, thus offering a reconfigurable platform for diverse microwave photonic computing functions. By programming the waveshaper to shape the comb lines according to the corresponding tap coefficients, this scheme can also apply to other computing functions such as Hilbert transforms[123]. The operation bandwidth can also be changed by adjusting the time delay between adjacent taps or employing different tap coefficients. An increased operation bandwidth can be achieved by simply employing a dispersive fibre with shorter length. The operation bandwidth is fundamentally limited by the Nyquist zone, which is determined by the comb spacing. In our case, the frequency spacing of the Kerr comb generated by the nonlinear MRR reaches 200 GHz, thus leading to a potential operation bandwidth of over 100 GHz.

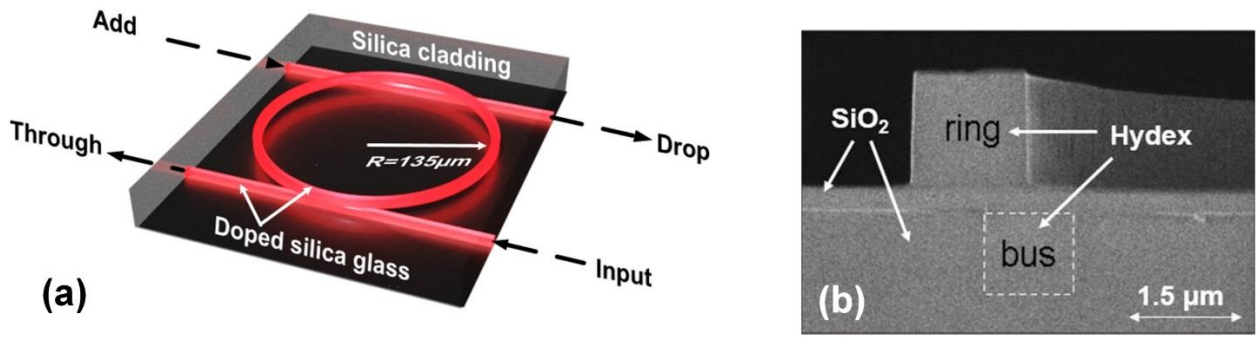

Fig. 3. (a) Schematic illustration of MRR. (b) SEM image of the cross-section of the MRR before depositing the $SiO_2$ upper cladding.

## 3. EXPERIMENTAL RESULTS

In our experiment, as Fig. 3(a) shows, the nonlinear MRR used to generate the Kerr comb was fabricated on a high-index doped silica glass platform using CMOS compatible fabrication processes [28]. First, high-index (n = 1.7) doped silica glass films were deposited using standard plasma enhanced chemical vapour deposition (PECVD), then photolithography and reactive ion etching (RIE) were employed to form waveguides with exceptionally low surface roughness. Finally, the waveguides were buried in a lower index upper cladding. The radius of the Hydex MRR is ~135 μm. The compact integrated MRR has a large free spectral range (FSR) of ~1.6 nm, i.e., ~200 GHz. Such a large FSR enables an increased Nyquist zone of ~100 GHz, which is challenging for mode-locked lasers and externally-modulated comb sources. The advantages of the platform for integrated nonlinear optics include ultra-low linear loss (~0.06 dB·$cm^{-1}$), a moderate nonlinearity parameter (~233 $W^{-1}·km^{-1}$), and in particular a negligible nonlinear loss up to extremely high intensities (~25 GW·$cm^{-2}$). After packaging the input and output ports of the device with fibre pigtails, the total insertion loss is ~3.5 dB. A scanning electron microscope (SEM) image of the cross-section of the MRR before depositing the $SiO_2$ upper cladding is shown in Fig. 3(b). By boosting the power of the CW light from the TLS via an EDFA and adjusting the polarization state, multiple mode-spaced combs were first generated, in which the primary spacing was determined by the parametric gain. When the parametric gain lobes became broad enough, secondary comb lines with a spacing equal to the FSR of the MRR were generated via either degenerate or non-degenerate four wave mixing (FWM). In our experiment, the power threshold for the generation of secondary comb lines was ~500 mW.

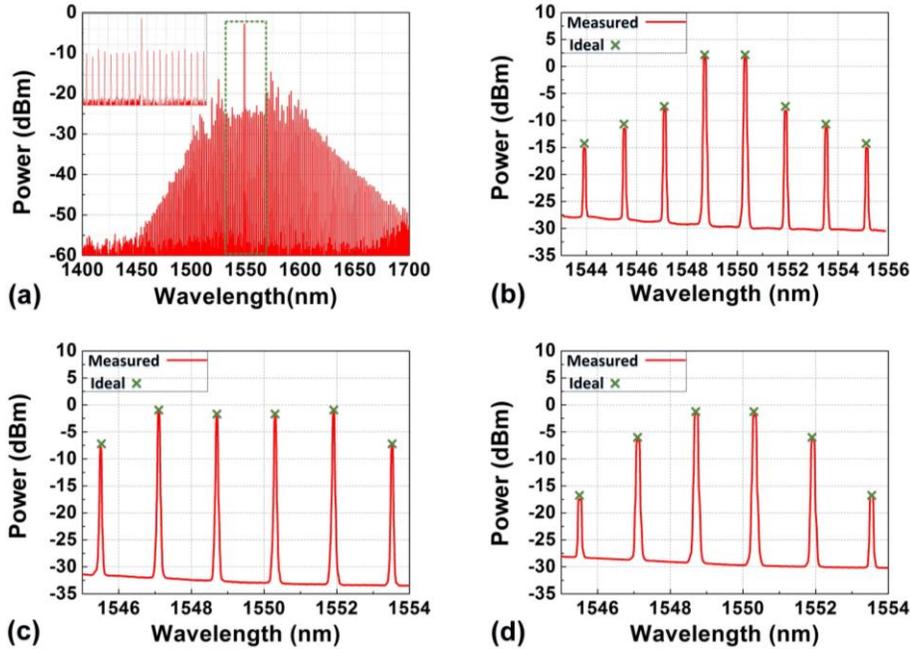

Fig. 4. (a) Optical spectrum of the generated Kerr comb in a 300-nm wavelength range. Inset shows a zoom-in spectrum with a span of ~32 nm. (b)–(d) Measured optical spectra (red solid) of the shaped optical combs and ideal tap weights (green crossing) for the first-, second-, and third-order intensity differentiators.

The resulting Type II Kerr optical comb [37] as shown in Fig. 4(a) is over 200-nm wide, and flat over ~32 nm. Since the generated comb only served as a multi-wavelength source for the subsequent transversal filter in which the optical power from different taps was detected incoherently by the photo-detector, the coherence of the comb was not crucial and the proposed differentiator was able to work under relatively incoherent conditions, although by necessity avoiding the chaotic regime [37]. In the experiment, the numbers of taps used for fractional-, first-, second-, and third-order differentiation demonstrations were 7, 8, 6, and 6, respectively. The choice of these numbers was made mainly by considering the power dynamic range of the comb lines, i.e., the difference between the maximum power of the generated comb lines and the power associated with the noise floor. The power dynamic range was determined by the EDFA before waveshaping, and in our case, it was ~30 dB. An increased number of taps requires a broader power dynamic range, which can be achieved by using an EDFA with a lower noise floor. As analysed in section 2, more taps are needed when the differentiation order increases, and for a fixed number of taps, increasing the order of differentiation also increases the required power dynamic range. In order to get better performance with a limited number of taps, we decreased the operation bandwidth of the second- and third-order differentiators to half of the transversal filter's Nyquist frequency when designing the response function with the Remez algorithm. It should be noted that the actual bandwidth of the differentiator is not limited by this design since the FSR of the transversal filter can be enlarged. The calculated tap coefficients for fractional-, first-, second-, and third-order differentiations are listed in Table I. The selected comb lines of the generated optical comb were processed by the waveshaper based on these coefficients. Considering that the generated Kerr comb was not flat, we adopted a real-time feedback control path to increase the accuracy of comb shaping. The comb lines' power was first detected by an optical spectrum analyzer (OSA) and compared with the ideal tap weights, and then an error signal was generated and fed back into the waveshaper to calibrate the system and achieve accurate comb processing. The shaped optical combs are shown in Figs. 4(b)–(d). A good match between the measured comb lines' power (red solid line) and the calculated ideal tap weights (green crossing) was achieved, indicating that the comb lines were successfully shaped. They were then divided into two parts according to the algebraic sign of the tap coefficients and fed into the 2×2 balanced MZM biased at quadrature. The modulated signal after the MZM was propagated through ~2.122-km single mode (dispersive) fibre (SMF). The dispersion of the SMF is ~17.4 ps/(nm·km), which corresponds to a time delay of ~59 ps between adjacent taps and yields an effective FSR of ~16.9 GHz in the RF response spectra.

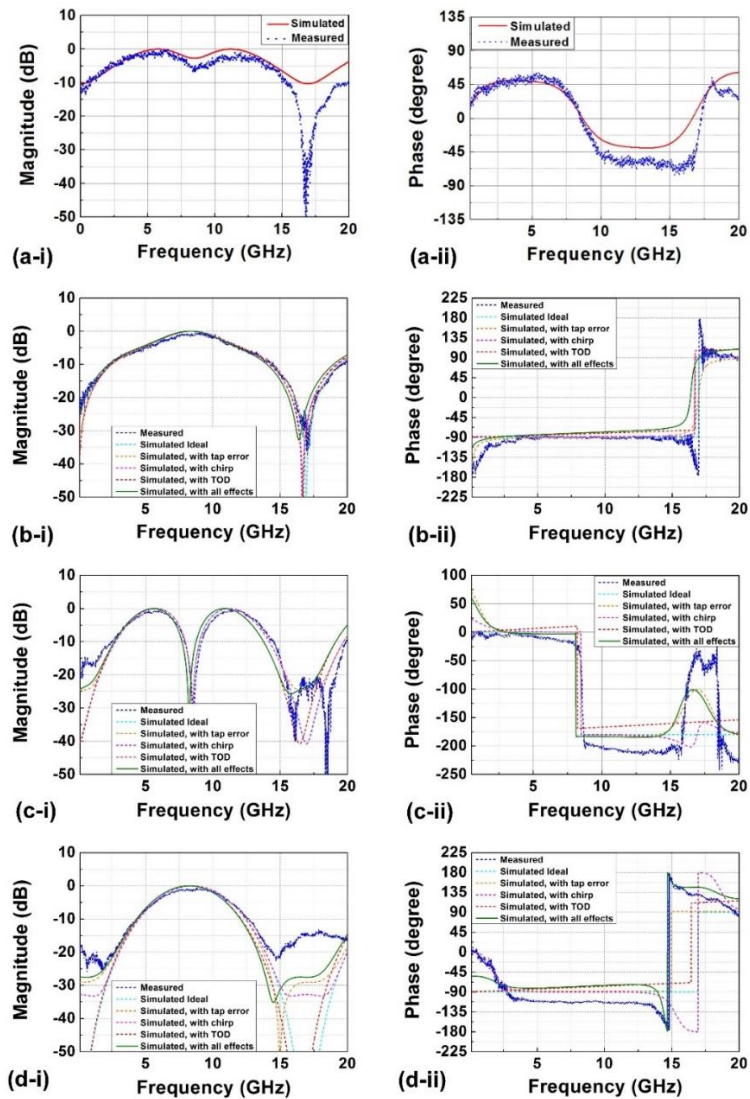

Fig. 5. Measured and simulated RF amplitude and phase responses of (a-i)–(a-ii) the fractional-order, (b-i)–(b-ii) the first-order, (c-i)–(c-ii) second-order, and (d-i)–(d-ii) third-order intensity differentiators. The simulated amplitude and phase responses after incorporating the tap error, chirp, and the third-order dispersion (TOD) are also shown accordingly.

## Table 1.

Tap coefficients for the fractional-, first-, second-, and third-order differentiations.

| Order of differentiation | Number of taps | Tap coefficients |
|---|---|---|
| Fractional-order | 7 | [0.0578, -0.1567, 0.3834, 1, -0.8288, 0.0985, -0.0892] |
| First-order | 8 | [−0.0226, 0.0523, −0.1152, 1, −1, 0.1152, −0.052, 0.0226] |
| Second-order | 6 | [0.0241, −0.1107, 0.0881, 0.0881, −0.1107, 0.0241] |

After the weighted and delayed taps were combined upon detection, the RF responses for different differentiation orders were characterized by a vector network analyser. Figures. 5(a-i), (b-i), (c-i) and (d-i) show the measured and simulated amplitude frequency responses of the first-, second-, and third-order intensity differentiators, respectively. The corresponding phase responses are depicted in Figs. 5(a-ii), (b-ii), (c-ii) and (d-ii). It can be seen that all three differentiators feature responses close to that expected from ideal differentiations. The FSR of the RF response spectra is ~16.9 GHz, which is consistent with the time delay between adjacent taps, as calculated before measurement. Note that by adjusting the FSR of the proposed transversal filter through the dispersive fibre or by programming the tap coefficients, a variable operation bandwidth for the intensity differentiator can be achieved, which is advantageous for meeting diverse requirements in operation bandwidth.

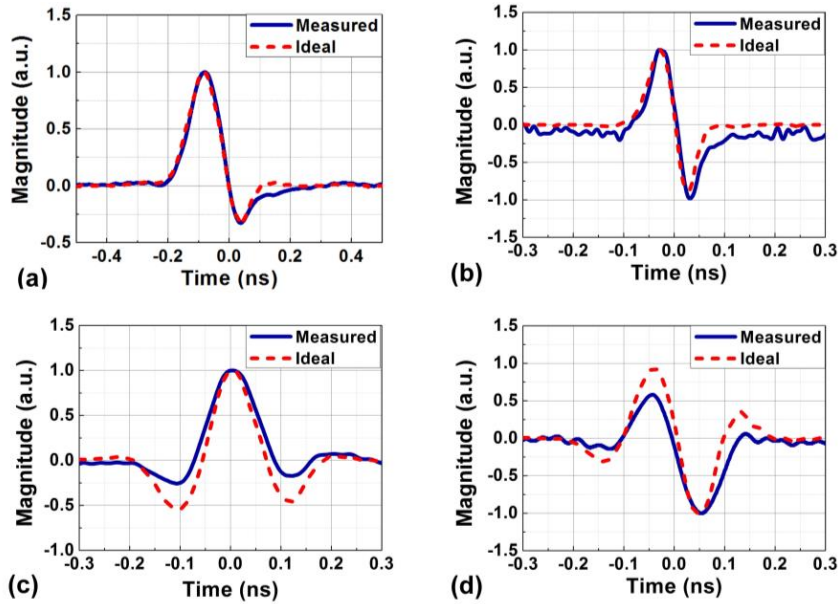

Fig. 6. Theoretical (red dashed) and experimental (blue solid) responses of the (a) fractional- (b) first-, (b) second-, and (c) third-order intensity differentiators.

We also performed [124] system demonstrations of real-time signal differentiations for Gaussian input pulses with a full-width at half maximum (FWHM) of ~0.12 ns, generated by an arbitrary waveform generator (AWG, KEYSIGHT M9505A). The waveform of the output signals after differentiation are shown in Figs. 6(a)–(d) (blue solid curves). They were recorded by means of a high-speed real-time oscilloscope (KEYSIGHT DSOZ504A Infiniium). For comparison, we also depict the ideal differentiation results, as shown in Figs. 6(a)–(d) (red dashed curves). The practical Gaussian pulse was used as the input RF signal for the simulation. One can see that the measured curves closely match their theoretical counterparts, indicating good agreement between experimental results and theory. Unlike the field differentiators, temporal derivatives of intensity profiles can be observed, indicating that intensity differentiation was successfully achieved. For the first-, second-, and third-order differentiators, the calculated RMSE between the measured and the theoretical curves are ~4.15%, ~6.38%, and ~7.24%, respectively.

# 4. RF PHOTONIC TRUE TIME DELAY LINES (TTDLS)

In modern radar and communications systems, RF photonic true time delay lines (TTDLs), which introduce multiple progressive time delays by means of RF photonics technologies, are basic building blocks with wide applications in phased array antennas (PAAs), RF photonic filters, analog-to-digital or digital-to-analog conversion, and arbitrary waveform generation [110, 130, 131]. For RF photonic TTDLs, increasing the number of delay channels can lead to significantly improved performance. For instance, in PAAs, the number of radiating elements determines the beamwidth, and so an improved angular resolution can be achieved by increasing the channel number [132]. Consequently, RF photonic TTDLs with a large number of channels are highly desirable for PAAs with high angular resolution. In conventional methods, discrete lasers arrays [110, 133] or fiber Bragg grating (FBG) arrays [132] are employed for implementing multiple TTDL channels, and so system cost and complexity significantly increase with the rising channel number, thus greatly limiting the number of available channels in practical systems. In contrast, for a broadband Kerr optical micro-comb generated by a single micro-resonator, a large number of high-quality wavelength channels can be simultaneously provided for the RF photonic TTDL, thus greatly reducing the size, cost, and complexity of the system [134].

Figure 7(a) shows a schematic diagram of a multi-channel RF photonic TTDL with an optical micro-comb source. The TTDL is constructed with three modules. The first module generates the optical micro-comb with a large number of wavelength channels using a micro-resonator. The second module shapes the optical micro-comb and directs it to a Mach-Zehnder modulator (MZM), where replicas of the input RF signal are generated on each wavelength to establish multiple RF channels. Time delays were then introduced between the RF channels by dispersive elements in the third module. The measured time delay responses of a 21-channel RF photonic TTDL based on the optical micro-comb generated by a Hydex MRR with a FSR of ~200 GHz are shown in Fig. 7(b). We used two different lengths of dispersive single mode fibres (SMFs), yielding two different delay steps of ~118 ps/channel and ~59 ps/channel. As shown in Fig. 7(c), we also measured the time delay responses of an 81-channel RF photonic TTDL based on the optical micro-comb generated by another Hydex MRR with a FSR of ~49 GHz and got two delay steps of ~29.6 ps/channel and ~14.8 ps/channel for different lengths of the dispersive SMF. The time delay steps were jointly determined by the frequency spacing of the generated optical comb and the accumulated dispersion of the SMF, yielding the four different delay steps shown in Figs. 7(b) and (c). The optical spectra of the Kerr combs generated by the Hydex MRRs are shown in Figs. 7(d-i) and (d-ii) accordingly.

One of the important applications of an RF photonic TTDL is in phased array antenna (PAA) systems, where optical signals on selected channels are separately converted into the electrical domain and then fed to an antenna array to form radiating elements. Considering that the antenna array is a uniformly spaced linear array with an element spacing of $d_{PAA}$, the steering angle $\theta_0$ of the PAA can be given as [111]

$$\theta_0 = \sin^{-1}\frac{c \cdot \tau}{d_{PAA}}, \tag{3}$$

where $c$ is the speed of light in vacuum, and $\tau = mT$ ($m = 1, 2, 3, \ldots$) is the time delay difference between adjacent radiating elements, with $T$ denoting the time delay difference between adjacent delay channels. The steering angle can be tuned by adjusting $\tau$, i.e., by changing the length of the dispersive medium or simply selecting every $m_{th}$ channel as radiating elements. The corresponding array factor $AF$ of the PAA can be expressed as [111]

$$AF(\theta, \lambda) = \frac{\sin^2[M\pi(d_{PAA}/\lambda)(\sin\theta - c \cdot mT/d_{PAA})]}{M^2 \sin^2[\pi(d_{PAA}/\lambda)(\sin\theta - c \cdot mT/d_{PAA})]} \tag{4}$$

where $\theta$ is the radiation angle, $M$ is the number of radiating elements, and $\lambda$ is the wavelength of the RF signals. The angular resolution of the PAA is the minimum angular separation at which two equal targets at the same range can be separated. It is determined by the 3-dB beamwidth of the PAA, which can be approximated as $\theta_{3dB} = 102/M$ [112], thus indicating that the angular resolution would greatly increase with a larger number of radiating elements, and this can be realised via optical micro-combs that can provide numerous wavelength channels for beam steering.

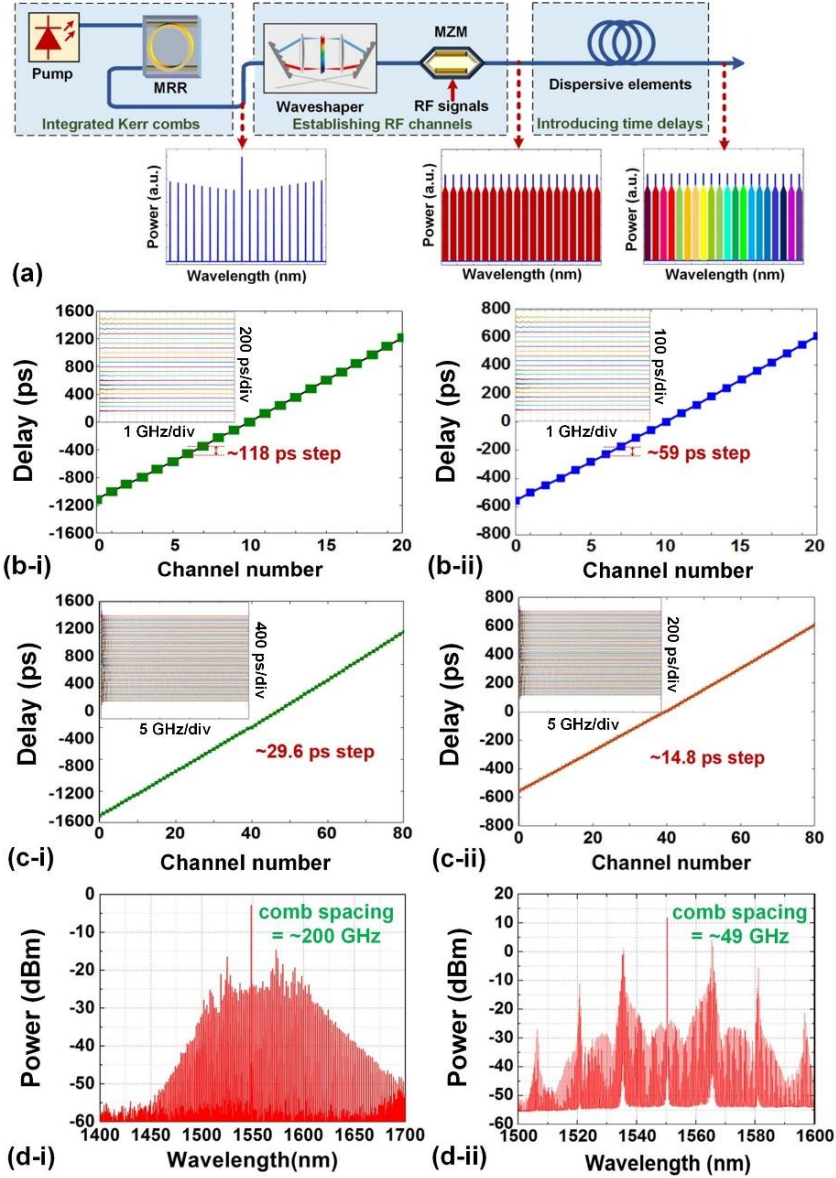

Fig. 7. RF photonic TTDLs based on Hydex optical micro-comb sources. (a) Schematic diagram. MRR: micro-ring resonator. MZM: Mach-Zehnder modulator. (b) Measured time delay responses of a 21-channel TTDL with (i) 4-km-long dispersive SMF and (ii) 2-km-long dispersive SMF. (c) Measured time delay responses of an 81-channel TTDL with (i) 4-km-long dispersive SMF and (ii) 2-km-long dispersive SMF. Insets in (b) and (c) show the flat delays over a wide RF frequency range. (d) Optical spectra of the micro-combs generated by (i) the Hydex MRR with a FSR of ~200 GHz and (ii) the Hydex MRR with a FSR of ~49 GHz.

Figure 8(a) depicts the relationship between the number of radiating elements $M$ and the 3-dB beamwidth $\theta_{3dB}$ of the PAA, where smaller beamwidths can be obtained when $M$ increases, therefore suggesting, given the large number of available channels provided by the optical micro-comb based RF photonic TTDL, that the angular resolution for the PAA can be greatly enhanced. The improved angular resolution for increased $M$ can also be observed in Fig. 8(b), which shows the $AF$ of the PAA for different $M$ calculated based on Eq. (4). To achieve a tunable beam steering angle, we selected every $m_{th}$ ($m$ = 1, 2, 3, …) wavelength of the TTDL by using the waveshaper, in such a way that the time delay ($\tau$) between the radiating elements could be varied with a step size of $T$. Figures. 8(c) and (d) show the calculated $AF$ for $m$ ($m$ = Channel number / $M$) varying from 1 to 7 based on a 200-GHz-FSR Hydex optical micro-comb and varying from 1 to 27 based on a 49-GHz-FSR Hydex optical micro-comb, respectively. As can be seen, the 49-GHz-FSR Kerr comb enables a much larger $M$ as $m$ varies, thus leading to a much smaller beamwidth and greatly enhanced angular resolution, as well as finer tuning steps and a large tuning range of the beam steering angle because of the larger number of channels (*m*). Moreover, PAAs based on RF photonic TTDLs can also achieve wide instantaneous bandwidths without beam squint (the variation of beam steering angle as a function of RF frequency). As indicated in Figs. 8(e) and (f), the beam steering angle remains the same while the RF frequency varies. These results confirm that PAAs based on RF photonic TTDLs with an optical micro-comb source feature greatly improved angular resolution, a wide tunable range in terms of beam steering angle, and a large instantaneous bandwidth. By using tunable dispersive elements to replace the SMF, allowing a tunable adjustment in time delay [136], optical micro-comb based two-dimensional PAAs with continuously-tunable time delays can be further realised.

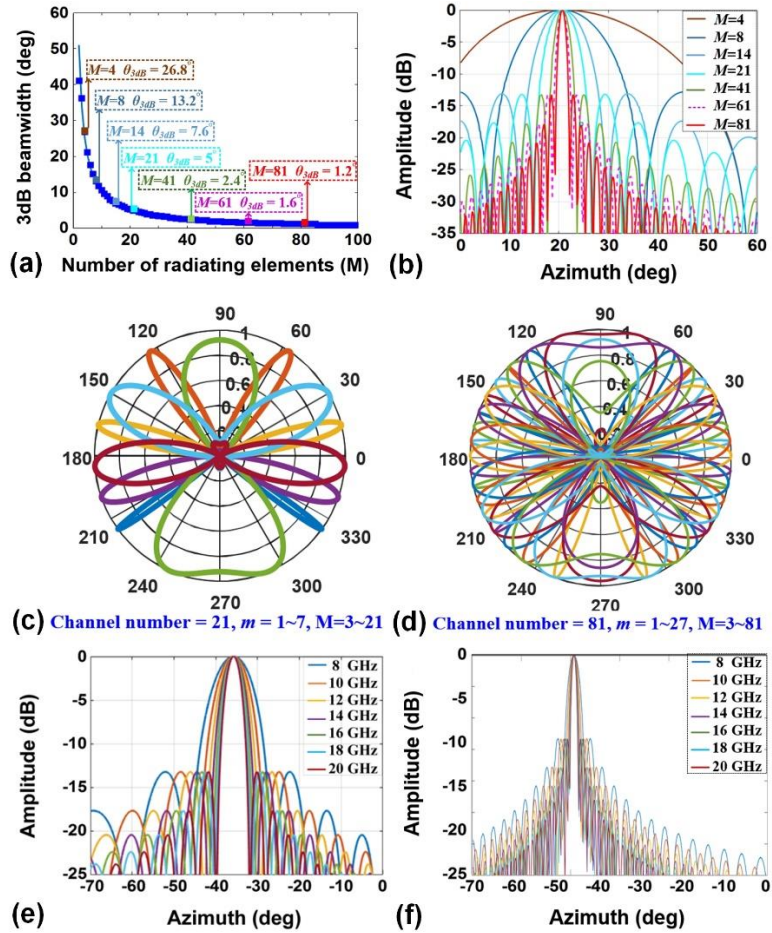

Fig. 8. (a) Relationship between the number of radiating elements *M* and the 3dB beamwidth $\theta_{3dB}$ of the PAA. (b) Calculated *AF* of the PAA with *M* varying from 4 to 81 based on a 49-GHz-FSR Hydex optical micro-comb. (c) Calculated *AF* of the PAA with *m* varying from 1 to 7 based on a 200-GHz-FSR Hydex optical micro-comb. (d) Calculated *AF* of the PAA with *m* varying from 1 to 27 based on a 49-GHz-FSR Hydex micro-comb. (e) Calculated *AF* of the PAA with various RF frequencies based on a 200-GHz-FSR Hydex optical micro-comb. (f) Calculated *AF* of the PAA with various RF frequencies based on a 49-GHz-FSR Hydex optical micro-comb.

## 5. CONCLUSION

Optical micro-combs represent an exceptionally active field of research that has only developed in the last ten years. This young and fast-growing field has its roots in the pioneering development of high-Q micro-cavity technologies, starting in the late 1980s and experiencing significant recent advances, both for bulk and integrated resonators. We review our recent progress in using optical micro-combs in the context of RF and microwave photonic signal processing. We demonstrate a reconfigurable microwave photonic intensity differentiator based on an integrated Kerr comb source. The Kerr optical comb is produced via a CMOS-compatible nonlinear MRR, which greatly increases the processing bandwidth and has the potential for a low cost, size, and complexity processing system. By programming and shaping the individual comb lines' power according to the calculated tap weights, we successfully demonstrate the first-, second-, and third-order intensity differentiations of the RF signal. The RF amplitude and phase responses of the proposed differentiator are characterized, and system demonstrations of real-time differentiations are performed for Gaussian input pulses. We achieve good agreement between theory and experimental results, thus verifying the effectiveness of our method. We also demonstrate optical true time delays for advanced RF radar applications. Our approach of using microcombs for RF and microwave signal processing provides a new way to implement microwave photonic signal processing functions with compact device footprint, high processing bandwidth, and high reconfigurability, thus holding great promise for future ultra-high-speed computing and information processing.

## 6. ACKNOWLEDGMENTS

This work was supported by the Australian Research Council Discovery Projects Program (No. DP150104327). RM acknowledges support by Natural Sciences and Engineering Research Council of Canada (NSERC) through the Strategic, Discovery and Acceleration Grants Schemes, by the MESI PSR-SIIRI Initiative in Quebec, and by the Canada Research Chair Program. He also acknowledges additional support by the Government of the Russian Federation through the ITMO Fellowship and Professorship Program (grant 074-U 01) by the 1000 Talents Sichuan Program. BEL was supported by the Strategic Priority Research Program of the Chinese Academy of Sciences, Grant No. XDB24030000.